\documentclass[%
 preprint,
 amsmath,amssymb,
 aps,
]{revtex4-2}

\usepackage{graphicx}
\usepackage{color}

\usepackage{dcolumn}
\usepackage{bm}

\begin{document}

\preprint{}

\title{Adaptive obstacle deposition drives dynamical localization in chase-and-escape dynamics}

\author{R. G. Rossatto}
 \affiliation{Department of Physics, Institute of Physics and Mathematics, Federal University of Pelotas, Pelotas, Brazil.}
\author{Amin Bakhshandeh}%
\affiliation{School of Chemistry and Food, Federal University of Rio Grande, Rio Grande 96203-900, Brazil.
}%

\author{H. Ariel Alvarez}
\affiliation{Institute of Physics of Liquids and Biological Systems, CONICET--UNLP, La Plata, Argentina.}%
\affiliation{
Institute of Health Sciences, Arturo Jauretche National University, Florencio Varela, Argentina}%
\affiliation{
Department of Biological Sciences, Faculty of Exact Sciences, National University of La Plata, La Plata, Argentina}%

\author{C. Manuel Carlevaro}
\affiliation{Institute of Physics of Liquids and Biological Systems, CONICET--UNLP, La Plata, Argentina.}%

\affiliation{%
National Technological University, La Plata Regional Faculty, Center for Experimental and Computational Mechanics Research, Berisso, Argentina}%

\author{José Rafael Bordin}
 \affiliation{Department of Physics, Institute of Physics and Mathematics, Federal University of Pelotas, Pelotas, Brazil.}
 \affiliation{Department of Physics, University of Konstanz, 78457 Konstanz, Germany.}
 \email{jrbordin@ufpel.edu.br}

\date{\today}

\begin{abstract}
Most pursuit--evasion models assume a fixed environment, although defensive actions can leave persistent structures that affect subsequent encounters. We study a two-dimensional lattice model in which threatened escapers deposit permanent obstacles, coupling capture kinetics to the geometry explored by their pursuers. Increasing the deposition probability enhances escaper survival and strongly suppresses chaser transport. For dilute chasers, the mean-squared-displacement exponent decreases from values near unity towards zero, trajectories contract to a few lattice spacings and the localized fraction rises sharply. The localization response becomes weaker as chaser abundance increases. Time-resolved obstacle densities support a kinetic explanation: rapid capture depletes obstacle-producing escapers before a comparably confining environment can develop. At intermediate parameters, broad or bimodal distributions of trajectory radii and MSD exponents reveal simultaneous mobile and localized subpopulations. Thus deposition redistributes trajectories between distinct dynamical behaviors, producing heterogeneity beyond a uniform slowdown. We summarize the coupled survival and transport response in an empirical map of mobile, heterogeneous and trapping-dominated regimes, interpreted as finite-time crossovers. These results identify the lifetime of the obstacle-producing population as a key control of adaptive localization: a stronger instantaneous pursuit pressure can weaken confinement by shortening the period over which the environment is modified.\end{abstract}

\maketitle


\section{Introduction}

Pursuit--evasion dynamics constitute a broad class of non-equilibrium processes in which one population attempts to approach, intercept or capture another population that simultaneously tries to avoid contact. Despite the apparent simplicity of this definition, local pursuit and escape rules can generate collective structures, coordinated strategies and non-trivial capture statistics that cannot be inferred directly from single-agent motion \cite{kamimura2010NJP,Angelani2012,Janosov2017,Ohira2024}. {\color{black}Extensions of the group chase-and-escape model have further shown that heterogeneity in chaser activity can improve capture \cite{masuko2017effect, somemori2025optimal}}. Such models provide a useful bridge between individual-level decisions and population-level outcomes because they preserve the directional asymmetry between chasers and escapers while remaining sufficiently minimal for systematic statistical analysis.

Defensive responses that leave persistent structures in the environment are common in nature. Members of the family Holothuriidae eject Cuvierian tubules when disturbed; the tubules elongate, adhere to nearby objects and can entangle a predator while the sea cucumber withdraws \cite{Flammang2002,Demeuldre2017}. Their regeneration is slow, so repeated threats progressively deplete this defence \cite{VandenSpiegel1987}. At a very different scale, protozoan grazing can stimulate microcolony formation and thickening of the exopolymeric matrix in bacterial biofilms, so that the medium through which a grazer moves is itself altered by predation pressure \cite{Matz2004,Weitere2005}. These examples motivate a general ingredient that is rarely included in minimal pursuit models: a local response to threat that persists after the encounter and constrains subsequent motion.

From the perspective of statistical physics, chase-and-escape systems belong to the broader family of active and driven many-body systems. Their microscopic rules violate detailed balance: chasers bias their motion towards targets, escapers bias their motion away from threats, and the resulting probability currents persist even in statistically stationary states. Persistence, exclusion, local crowding and interaction-induced correlations are known to generate anomalous transport and large dynamical fluctuations in active systems \cite{Vicsek2012,Bechinger2016PMR,Bowick22PRX,Epstein2019JCP,teVrugt2025EPJE}. Activity can also generate compact configurations through self-organized geometric constraints, as illustrated by chirality-induced self-wrapping in active polymers \cite{Caprini2025}. Pursuit introduces an additional source of non-equilibrium organization because the preferred direction of motion is continually updated by the evolving relative positions of the two populations. Capture efficiency is therefore controlled not only by the mobility of isolated agents, but also by how collective motion reshapes the set of accessible trajectories.

The physical environment is central to this problem. Walls, obstacles and bottlenecks constrain available paths and modify both the frequency and the geometry of encounters. This sensitivity appears across very different contexts. Immune and invasive cells migrate through dense, heterogeneous tissues whose architecture channels or impedes motion \cite{Fowell2021,Schienstock2021,MoreiraSoares2020,Melo2023}. Predator--prey encounters depend on habitat fragmentation, connectivity and barriers \cite{Gorini11,Cozzi2013}. Robotic pursuit and autonomous navigation require simultaneous target tracking, collision avoidance and path planning in obstacle-filled domains \cite{Oyler2016,Liang2023,Katona2024,Zhou2021,Alqudsi2025}, while aerial predators also alter pursuit trajectories to negotiate obstacles \cite{brighton2023CB}. Boundaries and congestion similarly affect escape routes and evacuation efficiency in pedestrian systems \cite{helbing2005TS,wang2025JCSS}.

Most theoretical treatments separate agent dynamics from environmental structure: the environment is generated first and then held fixed while the agents move through it. In lattice chase-and-escape models, fixed obstacles reduce accessible area, create detours and can divide the domain into weakly connected regions \cite{vscepanovic2019PhysicaA,rossatto2026confinement}. Active particles in heterogeneous media exhibit related subdiffusion, intermittent displacements and trapping \cite{chepizhko13,Bechinger2016PMR,Bellomo2020}. These studies establish that geometric disorder can strongly affect transport and capture, but they do not address the converse coupling in which pursuit itself generates the disorder that subsequently controls the dynamics.

This distinction between externally imposed and self-generated structure is important. In a quenched environment the obstacle field is statistically independent of the trajectories. In an adaptive environment, by contrast, obstacles are created at locations and times selected by the dynamics. The resulting disorder is correlated with threat events, escape routes and local population densities. The broader relevance of spatial correlations in disorder is illustrated by elastic networks, where correlated disorder modifies the packing fraction at the onset of rigidity \cite{EscobarAgudelo2026}. A trail left behind an escaper may close a corridor, redirect a chaser or combine with earlier deposits to form a local cage. Conversely, rapid capture terminates further modification of the environment. Geometry is therefore both an outcome of the pursuit and a feedback mechanism acting on its subsequent evolution.

Feedback between dynamical states and interaction structure is a defining feature of adaptive dynamical systems and adaptive networks \cite{berner2023PR,sawicki2023AIP}. When variables evolving on a network or substrate also alter that substrate, new time scales, heterogeneous states and sharp crossovers can emerge that are absent under fixed connectivity \cite{fialkowski2023PRL}. Related ideas have recently been explored for transport on adjustable networks, where the substrate responds to the moving particles \cite{oropesa2026transport}. Chase-and-escape dynamics provides a particularly transparent realization because environmental modification directly affects both the survival of the agents that create the structure and the transport of the agents pursuing them.

Here we introduce a minimal two-dimensional lattice model populated initially by escapers and chasers, in which a threatened escaper can transform the site it has just vacated into a permanent obstacle. The obstacle-deposition probability $P$ controls the strength of this environmental response, while the relative chaser fraction controls the frequency and duration of pursuit events. We show that this feedback couples obstacle accumulation, escaper survival and chaser localization, and produces a broad region of dynamical heterogeneity between predominantly mobile and trapping-dominated behavior.

\section{Model and methods}

\subsection{Dynamical rules and adaptive environment}

We consider a square lattice of linear size $L$ and total area $L^2$, with periodic boundary conditions in both directions. The full parameter scan uses $L=256$. Each lattice site may be empty or occupied by a single chaser, a single escaper or a static obstacle; multiple occupancy is prohibited. The initial escaper population corresponds to an areal fraction of approximately 0.10,
\begin{equation}
N_0^{E}\simeq 0.10L^2,
\end{equation}
and the number of chasers is
\begin{equation}
N^C=fN_0^E,
\qquad
f\in\{0.05,0.10,\ldots,1.00\}.
\end{equation}
The two populations are initially placed at random on distinct sites of an otherwise empty lattice. Chasers are conserved throughout a realization. Escapers are removed only through capture, and neither reproduction nor spontaneous death is included. Thus all persistent changes in the environment originate from the obstacle-deposition rule.

The dynamics is asynchronous and random sequential. At each elementary update, lattice sites are sampled uniformly until a site occupied by a mobile agent is found; the selected chaser or escaper then attempts one action. Empty and obstacle sites do not undergo independent updates. This is equivalent to choosing uniformly from the agents currently present. Because the escaper population decreases in time while $N^C$ remains fixed, the relative frequency with which chasers are selected increases as capture proceeds. We retain this microscopic update count as the time variable throughout the analysis rather than rescaling time by the instantaneous number of surviving agents.

Distances are evaluated using the Manhattan metric with the minimum-image convention. For two sites ${\bf r}_1=(x_1,y_1)$ and ${\bf r}_2=(x_2,y_2)$, we define
\begin{equation}
\Delta x=\min\bigl(|x_1-x_2|,L-|x_1-x_2|\bigr),
\qquad
\Delta y=\min\bigl(|y_1-y_2|,L-|y_1-y_2|\bigr),
\end{equation}
and
\begin{equation}
d_M({\bf r}_1,{\bf r}_2)=\Delta x+\Delta y.
\end{equation}
The capture range is one lattice spacing, whereas both species use sensing range $SR_C=SR_E=2$. Thus agents respond only to local information.

A selected chaser first inspects nearest-neighbour sites. If one or more escapers are found at $d_M=1$, one of them is chosen at random and captured, with the chaser moving to the target site. If no escaper is adjacent but at least one lies within $1<d_M\leq SR_C$, the chaser attempts a nearest-neighbour displacement in the direction associated with the largest local escaper population. If no escaper is detected, it attempts an unbiased nearest-neighbour step. Random selection resolves degeneracies between equivalent directions. The rule therefore interpolates between locally directed pursuit and an ordinary lattice random walk.

The escaper rule is the directional converse. If chasers are detected, the selected displacement favours motion away from the local threat, or towards the direction with the smallest local chaser population when threats occupy the outer part of the sensing region. In the absence of a detected chaser, the escaper attempts an unbiased nearest-neighbour step. Because sensing is local and multiple chasers may be present, the selected direction need not increase the distance from every chaser simultaneously. At large $f$, moving away from one threat can therefore bring an escaper closer to another.

\begin{figure}
    \centering
    \includegraphics[width=0.78\linewidth]{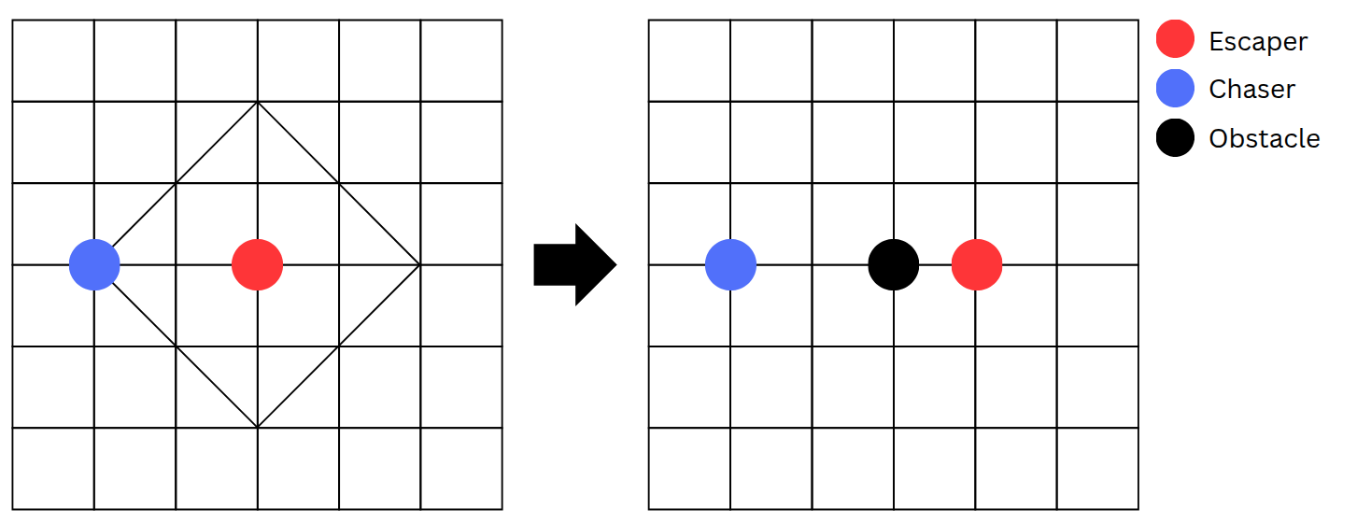}
    \caption{Schematic of adaptive obstacle deposition. A threatened escaper detects a chaser within its sensing range and moves away from the local threat. With probability $P$, the site vacated by the escaper is converted into a persistent obstacle. The schematic uses $P=1$ for clarity. Because deposition is conditioned on detection and successful escape, the resulting obstacle pattern is correlated with the pursuit history.}
    \label{fig:model}
\end{figure}

Proposed displacements obey excluded volume. Obstacles and agents occupying the destination site block the move, except when a chaser captures an adjacent escaper. When a threatened escaper successfully moves, its previous site is converted into a static obstacle with probability
\begin{equation}
P\in\{0,0.1,\ldots,1.0\}.
\end{equation}
The obstacle is permanent and impenetrable to both species. No obstacle is deposited after an unbiased move made without a detected chaser. Consequently, the obstacle field records a subset of threat-conditioned escaper displacements rather than the complete history of escaper motion. Figure~\ref{fig:model} illustrates one deposition event.

This rule generates two coupled feedback loops. Obstacle deposition can interrupt pursuit paths, reduce chaser mobility and prolong escaper survival. At the same time, captures reduce $N^E(t)$ and therefore remove the agents capable of creating further obstacles. At fixed $P$, a larger chaser population creates more potential threat events but also depletes the escaper population more rapidly; at fixed $f$, a larger $P$ increases the probability that a successful evasive displacement modifies the geometry. The observed dynamics reflects the competition between these two time scales.

\subsection{Simulation protocol and observables}

For every pair $(f,P)$ in the full $L=256$ scan, we perform 100 statistically independent realizations, each containing $10^7$ elementary agent updates. The initial positions and all stochastic choices are independently generated. Time in all time-dependent observables is reported in elementary updates. Summary curves are averages over independent realizations and error bars denote standard deviations; the latter are shown because the broad realization-to-realization and trajectory-to-trajectory fluctuations are themselves relevant near the heterogeneous regime. Selected comparisons at $L=128$, 256 and 512, together with tests of the localization threshold, are reported in the electronic supplementary material (ESM).

The normalized surviving escaper population is
\begin{equation}
S_E(t)=\frac{N^E(t)}{N_0^E},
\end{equation}
where $N^E(t)$ is the instantaneous number of escapers. By construction, $S_E(0)=1$ and $S_E(t)$ is non-increasing. The final value $S_E(t_{\rm max})$ measures the fraction that survives over the finite observation window and is therefore an operational survival probability rather than an asymptotic order parameter.

For each chaser $i$, the time-averaged mean-squared displacement is computed from its unwrapped stored trajectory. If the trajectory contains positions ${\bf r}_i(t_m)$, $m=1,\ldots,M$, the discrete estimator at lag $k$ is
\begin{equation}
\overline{\delta_i^2(k)}=
\frac{1}{M-k}\sum_{m=1}^{M-k}
\left|{\bf r}_i(t_{m+k})-{\bf r}_i(t_m)\right|^2.
\label{eq:tamsd}
\end{equation}
The corresponding lag time is the difference in elementary-update counts, $\Delta t=t_{m+k}-t_m$. Over the late-time range in which the log--log MSD is approximately linear, the trajectory is characterized by
\begin{equation}
\overline{\delta_i^2(\Delta t)}\sim \Delta t^{\alpha_i}.
\end{equation}
The same fitting procedure is applied to all parameter sets. The exponent $\alpha_i\simeq1$ indicates approximately diffusive motion, $\alpha_i<1$ subdiffusive motion and $\alpha_i>1$ superdiffusive motion. Values approaching zero signal saturation of the explored region on the simulated time scale. We report both $\langle\alpha\rangle$ and the distribution of individual exponents because the mean alone can obscure simultaneous mobile and nearly arrested trajectories.

The spatial extent of each chaser trajectory is measured independently through its temporal radius of gyration,
\begin{equation}
R_{g,i}^2=\frac{1}{M}\sum_{m=1}^{M}
\left|{\bf r}_i(t_m)-\overline{{\bf r}}_i\right|^2,
\qquad
\overline{{\bf r}}_i=\frac{1}{M}\sum_{m=1}^{M}{\bf r}_i(t_m).
\label{eq:rg}
\end{equation}
Here $R_g$ quantifies the spatial extension of successive positions visited by one trajectory rather than the size of a static many-particle configuration. $R_g$ and $\alpha$ are complementary: a small $R_g$ identifies a compact explored region, whereas a small $\alpha$ indicates slow growth of displacements over the fitting interval.

We classify a trajectory operationally as localized when
\begin{equation}
R_{g,i}\leq R_g^*=2,
\end{equation}
and define the localized fraction
\begin{equation}
\phi_T=\frac{N_T}{N^C},
\label{eq:phit}
\end{equation}
where $N_T$ is the number of chasers satisfying this criterion. The value $R_g^*=2$ is chosen from the structure of the measured $p(R_g)$ distributions: in the parameter interval where two dynamical populations are visible, it lies near the separation between the compact low-$R_g$ contribution and the broader mobile contribution. Thus the threshold is an operational classifier tied to the observed trajectory statistics rather than a thermodynamic order parameter.

The full distributions $p(R_g)$ and $p(\alpha)$ are retained in addition to their first moments. For the empirical regime classification below we use $R_g^*=2$ together with the operational exponent scale $\alpha_c=1/3$, which separates strongly suppressed transport from the broader subdiffusive and near-diffusive population in the measured distributions. Neither threshold is intended as a universal critical value. Trajectory observables were evaluated with TrajPy \cite{moreira2024trajpy}.

Finally, the realized obstacle density is
\begin{equation}
\rho_O(t)=\frac{N_O(t)}{L^2},
\end{equation}
where $N_O(t)$ is the number of persistent obstacles at time $t$. Unlike $P$, which is an input probability, $\rho_O(t)$ is a dynamical outcome and therefore depends on both the deposition probability and the lifetime of the escaper population.

\section{Results}

\begin{figure*}
    \centering
    \includegraphics[width=0.96\textwidth]{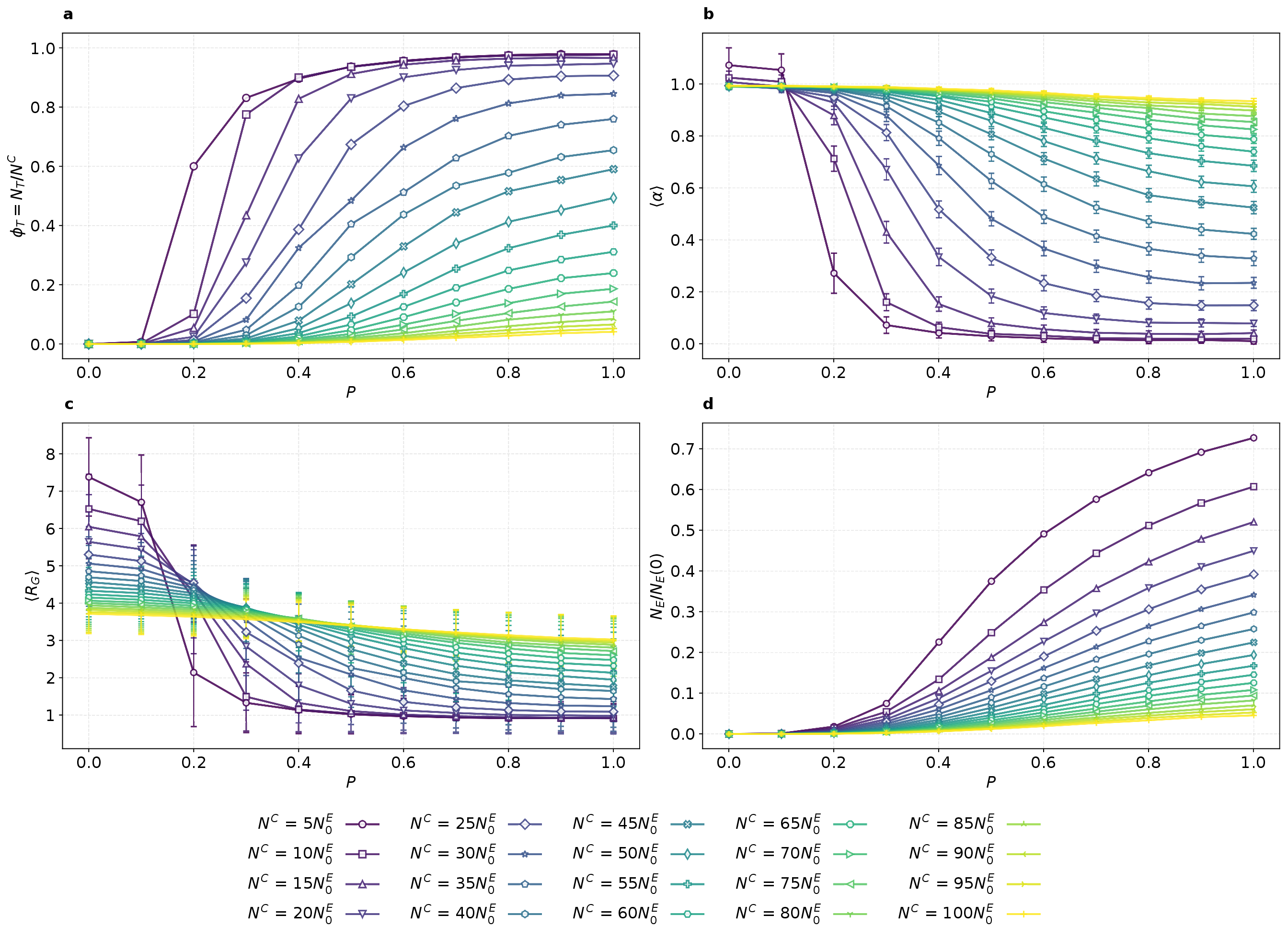}
    \caption{Coupled transport and survival response to adaptive obstacle deposition. (a) Fraction of localized chasers, defined by $R_g\leq2$. (b) Mean chaser MSD exponent. (c) Mean trajectory radius of gyration. (d) Final normalized escaper population. Curves span $f=0.05,0.10,\ldots,1.00$, as indicated in the legends. Error bars are standard deviations across independent realizations.}
    \label{fig:summary}
\end{figure*}

Figure~\ref{fig:summary} brings together the four observables that define the macroscopic response: the localized chaser fraction $\phi_T$, the mean MSD exponent $\langle\alpha\rangle$, the mean trajectory radius of gyration $\langle R_g\rangle$ and the final surviving escaper fraction. Their joint evolution is more informative than any individual curve. At $P=0$, no persistent environmental structure is generated and the chasers remain predominantly mobile. The most dilute systems exhibit the largest $R_g$ and weakly superdiffusive or nearly diffusive motion, consistent with long directed pursuit segments through an unobstructed target-rich environment. Increasing $f$ already reduces the spatial extent of individual trajectories through competition, excluded-volume blocking and faster target depletion, but it does not generate the strong localization observed when the environment can be modified.

The response to $P$ is strongly population dependent. For small $f$, $\phi_T$ rises sharply over a narrow interval of deposition probabilities, while $\langle\alpha\rangle$ simultaneously drops from values near unity towards strongly subdiffusive values and $\langle R_g\rangle$ collapses to a scale of only a few lattice spacings. As $f$ increases, all three changes become progressively weaker and shift towards larger $P$. The highest chaser fractions retain extended trajectories even when obstacle deposition is certain after every successful threat-conditioned escape. Thus adding more chasers does not reinforce trapping monotonically: a large pursuing population also removes the obstacle-producing population more rapidly.

Escaper survival shows the reciprocal dependence. In the absence of deposition, almost all escapers are captured within the observation window. Survival rises once adaptive obstacles become dynamically relevant and does so most strongly at small $f$. A dilute chaser population allows a long construction period during which repeated threat-conditioned moves can accumulate a protective obstacle geometry. A dense chaser population creates more immediate threats but shortens that construction period. The correlation between survival, $\phi_T$, $\alpha$ and $R_g$ therefore points to a kinetic feedback rather than to a simple reduction of the step acceptance rate.

{\color{black}The simultaneous and relatively sharp changes observed near \(P\simeq0.2\) for dilute chaser populations suggest that the localization crossover may represent a nonequilibrium dynamical transition. Determining whether the curves for different \(f\) admit a common scaling form will require systematic finite-size and finite-time analyses. Directed percolation is an intriguing possibility, although the role of connectivity in the self-generated obstacle field should also be examined. Finite-size comparisons for \(L=128\), \(256\), and \(512\), reported in the electronic supplementary material, reveal quantitative size effects but preserve the qualitative transport, localization, and survival trends discussed above.}

\begin{figure*}
    \centering
    \includegraphics[width=0.96\textwidth]{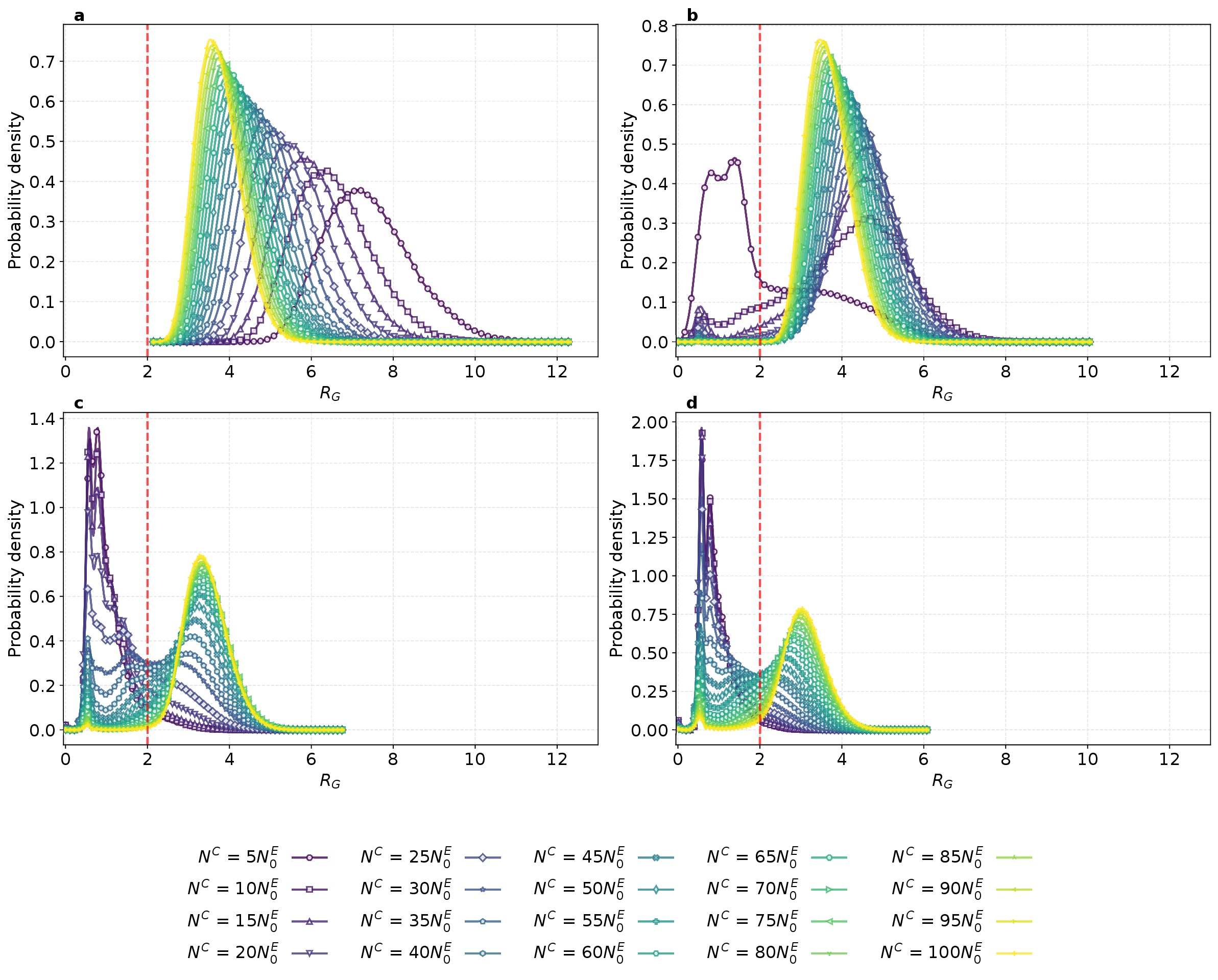}
    \caption{Representative distributions of the trajectory radius of gyration for (a) $P=0$, (b) $P=0.2$, (c) $P=0.5$ and (d) $P=1.0$. The dashed line marks the operational localization threshold $R_g^*=2$. Curves span $f=0.05,0.10,\ldots,1.00$, as indicated in the legends. Intermediate parameters display simultaneous low- and high-$R_g$ contributions, revealing strong dynamical heterogeneity.}
    \label{fig:rgdist}
\end{figure*}

Mean values alone cannot establish whether the crossover is homogeneous. Figure~\ref{fig:rgdist} shows representative distributions of the individual trajectory radius of gyration. At $P=0$, the distributions are predominantly unimodal and centred above the localization scale. At intermediate $P$, low-$f$ curves develop a distinct compact contribution below $R_g=2$ while retaining substantial weight at larger $R_g$. The same nominal $(f,P)$ therefore contains chasers exploring qualitatively different spatial environments. With increasing $P$, the low-$R_g$ contribution becomes more prominent for dilute and intermediate populations, whereas the highest-$f$ systems remain dominated by extended trajectories.

\begin{figure*}
    \centering
    \includegraphics[width=0.96\textwidth]{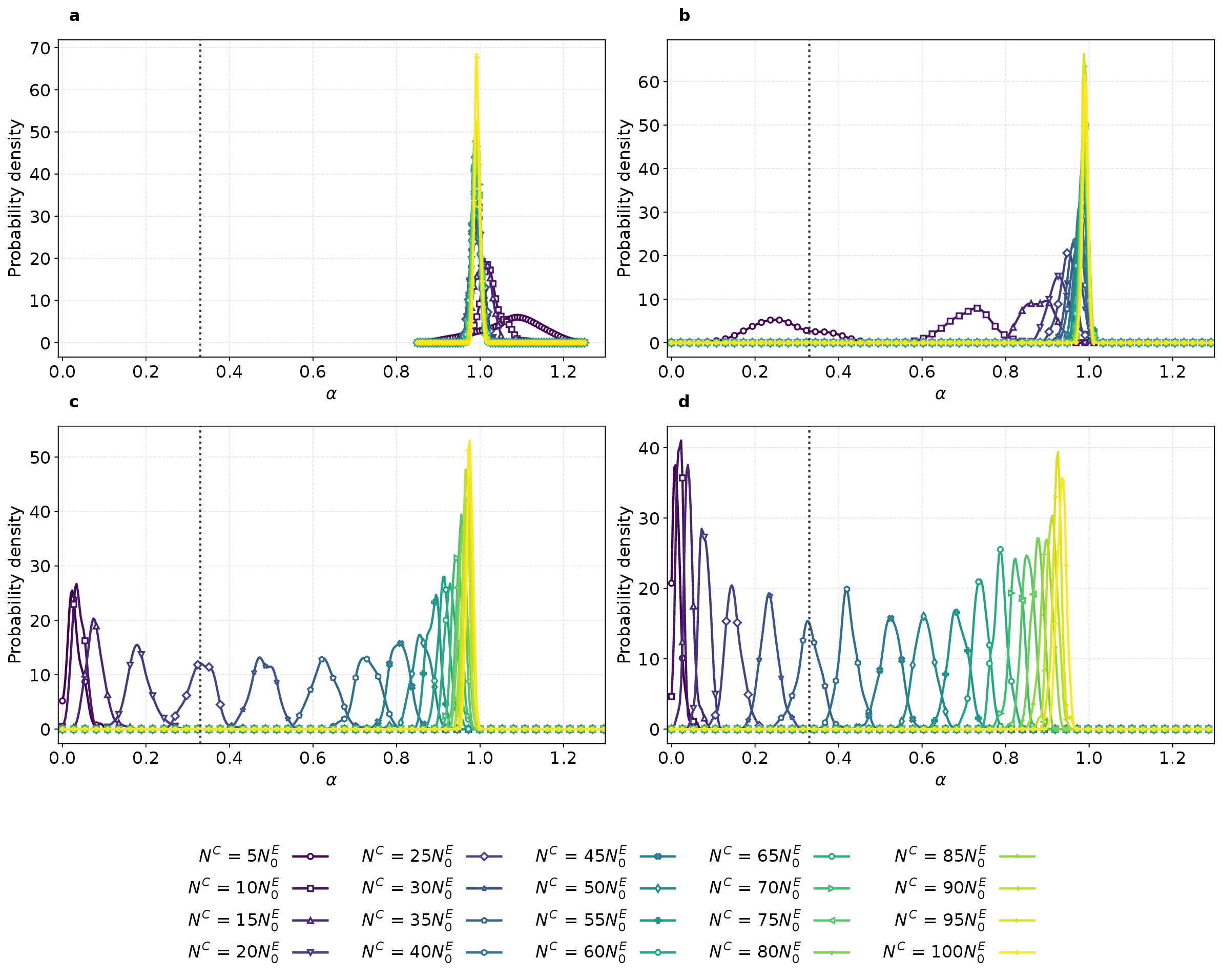}
    \caption{Representative distributions of fitted chaser MSD exponents for (a) $P=0$, (b) $P=0.2$, (c) $P=0.5$ and (d) $P=1.0$, matching the deposition probabilities in figure~\ref{fig:rgdist}. Curves span $f=0.05,0.10,\ldots,1.00$, as indicated in the shared legend. The dashed line marks the operational scale $\alpha_c=1/3$. Dilute chasers develop broad or multiple contributions at intermediate parameters and strongly suppressed exponents at high $P$, while dense populations retain near-diffusive motion. Exponents characterize the finite fitting interval.}
    \label{fig:alphadist}
\end{figure*}

The temporal signature of this heterogeneity is shown explicitly in figure~\ref{fig:alphadist}, using the same deposition probabilities as figure~\ref{fig:rgdist}. At $P=0$, most probability is concentrated near $\alpha=1$, with a weak superdiffusive contribution for dilute chasers. At intermediate $P$, some dilute-population distributions broaden or develop separated low- and high-$\alpha$ contributions, showing strongly suppressed and more mobile trajectories under the same global parameters. Increasing $P$ shifts the dilute-population distributions towards $\alpha=0$, whereas dense chaser populations retain substantial weight near unity even at $P=1$. The decrease of $\langle\alpha\rangle$ therefore includes a redistribution between trajectory classes, beyond a uniform reduction of mobility. The complete distributions for all sampled $P$ are provided in the ESM.

The broad distributions clarify why $\langle R_g\rangle$, $\phi_T$ and $p(\alpha)$ should be interpreted together. In the heterogeneous regime the mean radius can lie between two characteristic trajectory populations even when relatively few trajectories are actually represented by that intermediate value. Conversely, $\phi_T$ measures the weight below the chosen threshold but does not characterize the extent of the remaining mobile trajectories or their displacement growth. The spatial and temporal distributions provide complementary evidence for heterogeneous motion. We therefore use the term \emph{dynamical heterogeneity}, rather than phase coexistence: the data establish simultaneous mobile and localized trajectory classes in a finite system, but do not imply two thermodynamic phases.

{\color{black}This emergent separation between mobile and localized chasers is superficially reminiscent of the division of labour between diligent and lazy chasers considered in previous group chase-and-escape models \cite{masuko2017effect, somemori2025optimal}. The dynamical consequences, however, are different. In those models, randomly moving or partially lazy chasers retain sufficient spatial mobility to disrupt compact pursuit clusters and favour pincer-like configurations, thereby improving capture efficiency. Here, the heterogeneity is not prescribed at the agent level but emerges from the pursuit-conditioned obstacle field, and the low-mobility subpopulation becomes spatially confined. Consequently, the transfer of statistical weight towards small \(R_g\) and low MSD exponents is accompanied by increased escaper survival. This contrast suggests that reduced pursuit directedness may assist capture only while chasers retain enough mobility to reorganize the pursuit geometry; once persistent obstacles generate spatial localization, the loss of mobility dominates.}

Tests with $R_g^*=1$, 2 and 4 in the ESM show that the numerical localized fraction depends appreciably on the classifier. Dilute chasers display a strong deposition-induced increase for all three thresholds, but $R_g^*=4$ also classifies a substantial fraction of dense-population trajectories as compact at $P=0$, where their MSD exponents remain near unity. We therefore retain the distribution-based choice $R_g^*=2$ and interpret it jointly with $p(\alpha)$; spatial compactness over a finite trajectory is not sufficient by itself to establish arrest.

\begin{figure*}
    \centering
    \includegraphics[width=0.96\textwidth]{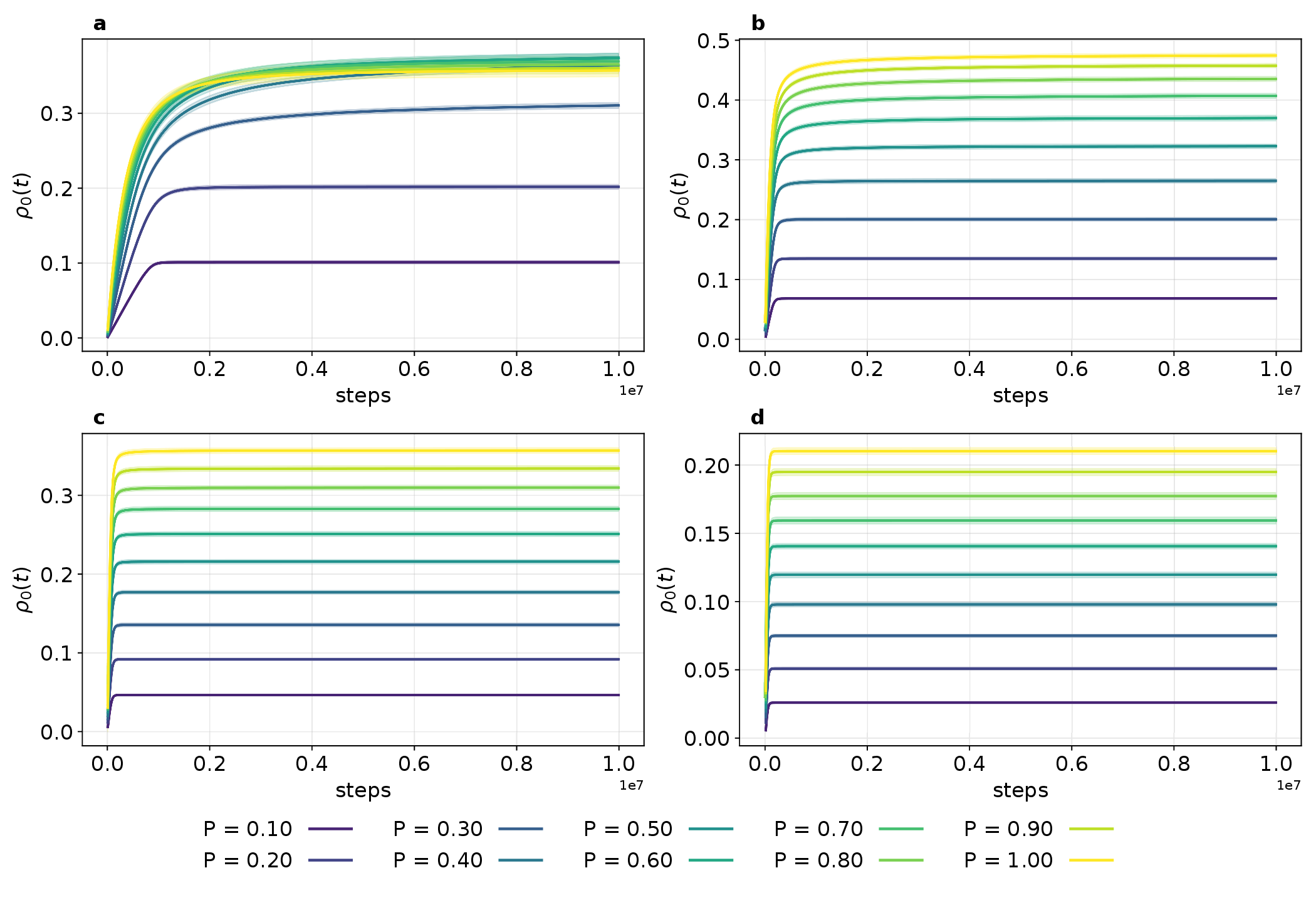}
    \caption{Time evolution of the realized obstacle density $\rho_O(t)=N_O(t)/L^2$ for (a) $f=0.05$, (b) $f=0.25$, (c) $f=0.50$ and (d) $f=1.00$. Curves correspond to deposition probabilities $P=0.1,0.2,\ldots,1.0$. Time is reported in elementary agent updates. Increasing $P$ accelerates obstacle accumulation, whereas increasing the chaser fraction eventually suppresses the final obstacle density by shortening the lifetime of the escaper population.}
    \label{fig:obstime}
\end{figure*}

The adaptive mechanism is seen directly in the accumulation of persistent obstacles. Figure~\ref{fig:obstime} shows $\rho_O(t)$ for four representative chaser fractions. For all populations, increasing $P$ increases the rate and final extent of obstacle deposition, but the realized density is not determined by $P$ alone. At low and intermediate $f$, pursuit persists for long enough that the obstacle field continues to grow before reaching a plateau. At $f=1$, by contrast, the curves saturate rapidly at substantially lower densities because escapers are captured before they can repeatedly modify the lattice. The obstacle field is therefore a dynamical record of the competition between deposition and depletion.

This behavior explains why $P$ should not be interpreted as an obstacle-density control parameter. The same deposition probability can generate very different environments because the number and spatial history of successful escape events depend on $f$. In particular, strong localization appears where pursuit persists long enough for repeated local modifications to reorganize accessible paths. The response is therefore consistent with a self-reinforcing mechanism: deposited obstacles protect some escapers and hinder nearby chasers, which prolongs the lifetime of the population capable of depositing additional obstacles. Capture provides the competing self-limiting process by removing those obstacle producers.

\begin{figure*}
    \centering
    \includegraphics[width=0.84\textwidth]{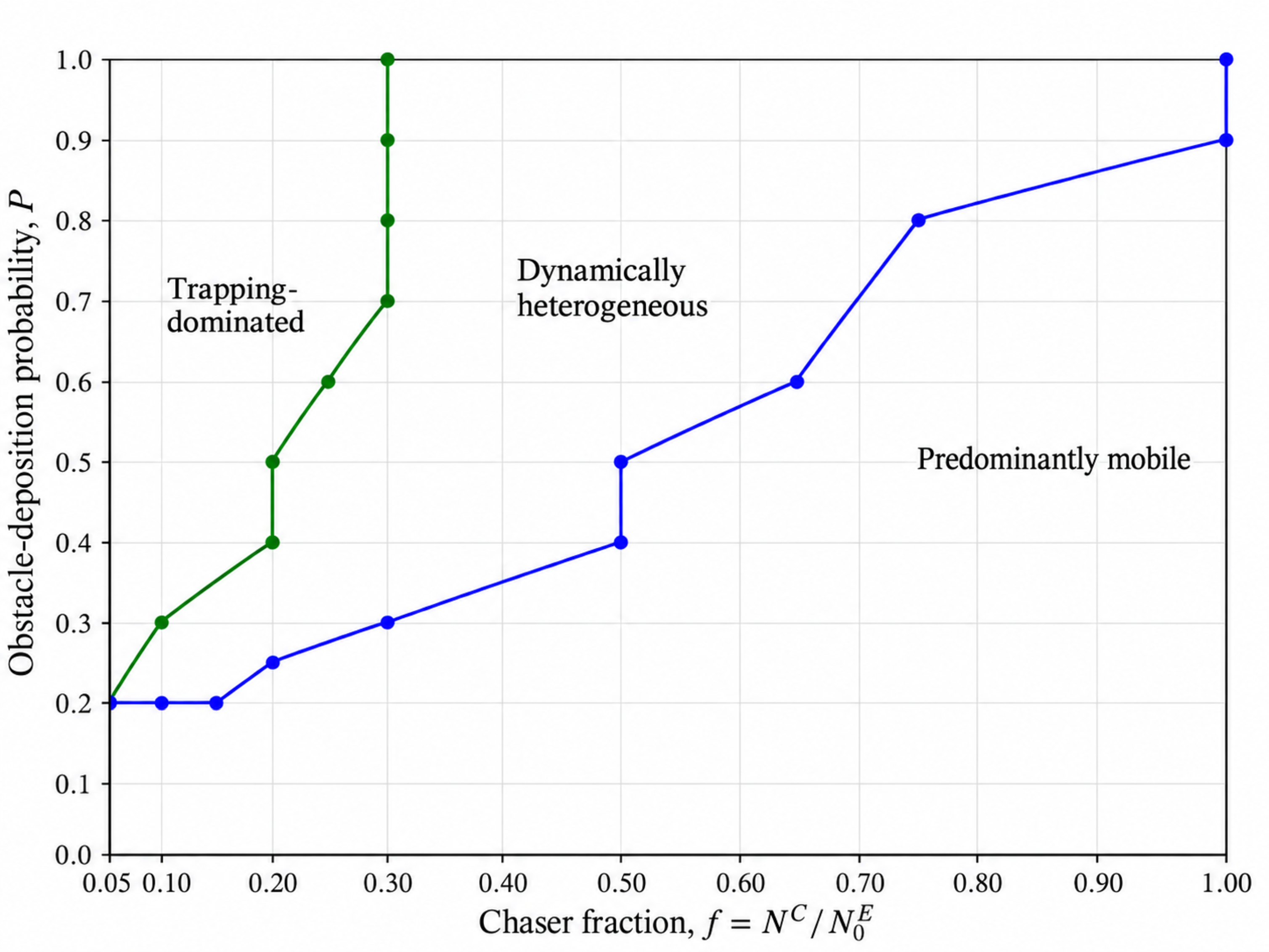}
    \caption{Empirical dynamical-regime map in the $(f,P)$ plane for $L=256$. The three regions denote predominantly mobile, dynamically heterogeneous and trapping-dominated behavior according to the operational trajectory criteria described in the text. Symbols mark crossover estimates on the sampled parameter grid and connecting lines are guides to the eye, not thermodynamic phase boundaries. {\color{black}The regime boundaries summarize the combined behavior of \(\phi_T\), \(\langle\alpha\rangle\), \(\langle R_g\rangle\), and the corresponding trajectory distributions, rather than a threshold in a single scalar observable.}}
    \label{fig:map}
\end{figure*}

The combined trajectory distributions can be summarized by an empirical dynamical-regime map (figure~\ref{fig:map}). We classify sampled $(f,P)$ points using the positions and shapes of the measured $p(R_g)$ and $p(\alpha)$ distributions. A state is termed predominantly mobile when the principal trajectory population lies above $R_g^*=2$ and on the weakly subdiffusive or near-diffusive side of $\alpha_c=1/3$. It is termed trapping-dominated when the principal low-$R_g$ population lies below $R_g^*$ and the corresponding exponent distribution is concentrated in the strongly suppressed regime. Parameter sets containing resolved contributions on both sides of these operational thresholds are classified as dynamically heterogeneous. Boundary lines are drawn between neighbouring sampled points assigned to different classes and are therefore guides to crossovers rather than fitted critical lines. Their resolution is limited by the parameter grid, $\Delta f=0.05$ and $\Delta P=0.1$.

The map emphasizes two robust trends. First, the onset of strongly suppressed transport shifts towards larger $P$ as $f$ increases. Second, the trapping-dominated region disappears from the explored interval at sufficiently large chaser fractions, where rapid target depletion prevents extensive adaptive restructuring. The size comparisons in the ESM preserve the contrast between dilute and dense populations, but reveal quantitative differences in all four summary observables. In particular, $\langle R_g\rangle$ decreases with increasing $L$ even at $P=0$, the high-$P$ localized fraction grows markedly with $L$ at $f=0.50$, and dilute chasers retain a larger residual MSD exponent in the largest lattice.

These checks support the persistence of the qualitative trends over the supplied observation windows, rather than convergence of the crossover boundaries. With time counted in elementary updates, a fixed total update budget gives fewer updates per agent as the initial populations grow with $L^2$. The comparisons therefore do not isolate spatial size effects from finite observation time per agent. The map remains an operational summary of the $L=256$ scan, with size, observation time and classifier choice contributing to its quantitative limitations.

The adaptive model also differs conceptually from our earlier chase-and-escape studies in quenched obstacle fields \cite{vscepanovic2019PhysicaA,rossatto2026confinement}. In both settings, excluded sites can reduce mobility and enhance survival. In the present model, however, obstacles are created specifically during successful evasive events and their number is dynamically coupled to capture. The resulting localization is therefore not controlled by obstacle density alone: the environment retains information about where pursuit occurred and about how long escapers survived to modify it. Establishing which geometric correlations are most effective---for example, local cages, dead ends or changes in accessible-space connectivity---is a natural direction for future work, but it is not required to identify the feedback-induced localization documented by the trajectory observables.

\section{Conclusion}

We have shown that a local defensive response can couple escaper survival and chaser transport through the progressive construction of an obstacle field. When threatened escapers deposit permanent obstacles after successful displacements, increasing the deposition probability enhances survival and suppresses chaser motion, particularly for dilute pursuing populations. This suppression is reflected jointly in the decrease of the fitted MSD exponents towards zero, the contraction of trajectories to a few lattice spacings and the increase of the operationally localized fraction. Together, these observations connect microscopic escape events to collective changes in survival and transport.

The dependence on chaser abundance identifies a competition between environmental construction and the depletion of the population responsible for it. A larger chaser population both creates more potential threats and accelerates the removal of obstacle-producing escapers. For dilute chasers, longer escaper lifetimes allow repeated deposits to accumulate and constrain subsequent pursuit. For dense chaser populations, rapid capture shortens this construction period, limiting obstacle accumulation even when deposition is certain after every successful threat-conditioned displacement. The obstacle-density histories support this kinetic interpretation. Thus, the deposition probability controls the likelihood of environmental modification per event, while capture kinetics helps determine how long that modification can continue. Stronger pursuit pressure can consequently weaken the confinement generated by the defensive response.

The distributions of trajectory radii and MSD exponents further show that the transport response involves substantial dynamical heterogeneity. At intermediate parameters, compact and extended trajectories occur under the same global conditions, accompanied by broad or multiple contributions to the exponent distributions. The evolution of the mean observables therefore includes changes in the relative statistical weights of distinct dynamical behaviors. The spatial distributions characterize the extent of the explored regions, while the exponent distributions characterize displacement growth over the fitting interval. Their complementary information reveals mobile and nearly arrested trajectory populations that would be obscured by an analysis restricted to averages.

The empirical regime map organizes these observations into predominantly mobile, dynamically heterogeneous and trapping-dominated behavior. Its boundaries describe operational crossovers over the simulated observation window. Comparisons across lattice sizes preserve the main contrast between dilute and dense chaser populations, while revealing quantitative changes in the transport and survival observables. Because equal total update budgets provide different observation times per agent as the system size changes, spatial and temporal effects remain coupled in these controls. Threshold tests likewise document the sensitivity of the numerical localized fraction. These findings support interpreting localization through the combined spatial and temporal trajectory statistics and delimit the quantitative scope of the map.

The central physical implication is that the lifetime of the obstacle-producing population helps control localization in an adaptive environment. The obstacle field retains a spatial record of successful escape events, while its accumulated extent reflects how long escapers survived to modify accessible paths. Capture therefore affects both the population balance and the subsequent environment through which pursuit occurs. Future work could quantify the geometric correlations responsible for confinement, including local cages and changes in accessible-space connectivity, and examine how finite obstacle lifetimes or escaper replenishment modify the feedback. More generally, the model shows how local responses to encounters can produce persistent constraints on motion through the coupled evolution of populations and environmental memory.

The model also provides a basis for exploring other forms of agent–environment coupling. Run-and-tumble or chiral dynamics could reveal how persistence and trajectory curvature affect localization, while slowly diffusing or pushable obstacles would interpolate between persistent and dynamically reorganized environments. A finite obstacle-deposition budget would introduce resource depletion, and extending the model to three dimensions would test whether increased path connectivity suppresses localization. An inverse remodeling rule, in which escapers clear a path through a pre-existing obstacle field, could instead facilitate pursuit by leaving an accessible wake. Related studies have shown that driven particles moving through responsive backgrounds can generate mediated effective interactions and collective transport, suggesting that such a rule could produce qualitatively new cooperative pursuit regimes \cite{Reichhardt2006,Vasilyev2017}.

\section*{Data accessibility}
The simulation code, analysis scripts and processed data required to reproduce all figures are available on GitHub:\\
\url{https://github.com/RafaRossatto/Chase-escape_with_temporary_obstacles}

\section*{Authors' contributions}
R.G.R.: software, formal analysis, data curation, validation, writing---original draft; A.B.: formal analysis, writing---review and editing; H.A.A., C.M.C. and J.R.B.: conceptualization, funding acquisition, project administration, methodology, resources, supervision, writing---review and editing.

\section*{Competing interests}
The authors declare no competing interests.

\section*{Funding}
This research was made possible by public funding. J.R.B. and A.B. acknowledge financial support from the Brazilian National Council for Scientific and Technological Development (CNPq), grants 309600/2026-0 and 317323/2025-3, and from the Funda\c{c}\~ao de Apoio \`a Pesquisa do Estado do Rio Grande do Sul (FAPERGS), grants 25/2551-0002609-1 and 26/2551-9000087-4. R.G.R. and J.R.B. acknowledge scholarship support from the Coordination for the Improvement of Higher Education Personnel (CAPES, Finance Code 001). J.R.B. also acknowledges Alexander von Humboldt Foundation for financial support through a research fellowship at the University of Konstanz. C.M.C. acknowledges financial support from CONICET (PIP 11220210100884CO) and Universidad Tecnol\'ogica Nacional (PID MATCLP10087C), while H.A.A. acknowledges financial support from CONICET (PIP 11220210100884CO) and Universidad Nacional Arturo Jauretche (UNAJ Investiga 80020230300017UJ).
\nocite{*}

\bibliography{apssamp}

@article{moreira2024trajpy,
  title={TrajPy: empowering feature engineering for trajectory analysis across domains},
  author={Moreira-Soares, Maur{\'\i}cio and Mossmann, Eduardo and Travasso, Rui DM and Bordin, Jos{\'e} Rafael},
  journal={Bioinformatics Advances},
  volume={4},
  number={1},
  pages={vbae026},
  year={2024},
  publisher={Oxford University Press},
  doi = {10.1093/bioadv/vbae026}
}

@article{ramon2023virulence,
  title={Virulence factors of Mycobacterium tuberculosis as modulators of cell death mechanisms},
  author={Ramon-Luing, Lucero A and Palacios, Yadira and Ruiz, Andy and T{\'e}llez-Navarrete, Norma A and Chavez-Galan, Leslie},
  journal={Pathogens},
  volume={12},
  number={6},
  pages={839},
  year={2023},
  publisher={MDPI},
  doi = {10.3390/pathogens12060839}
}

@article{kamimura2010NJP,
  title={Group chase and escape},
  author={Kamimura, Atsushi and Ohira, Toru},
  journal={New Journal of Physics},
  volume={12},
  number={5},
  pages={053013},
  year={2010},
  publisher={IOP Publishing},
  doi = {10.1088/1367-2630/12/5/053013}
}

@article{Angelani2012,
  title = {Collective Predation and Escape Strategies},
  volume = {109},
  ISSN = {1079-7114},
  url = {http://dx.doi.org/10.1103/PhysRevLett.109.118104},
  DOI = {10.1103/physrevlett.109.118104},
  number = {11},
  journal = {Physical Review Letters},
  publisher = {American Physical Society (APS)},
  author = {Angelani,  Luca},
  year = {2012},
  month = sep 
}

@article{Vicsek2012,
  title = {Collective motion},
  volume = {517},
  ISSN = {0370-1573},
  url = {http://dx.doi.org/10.1016/j.physrep.2012.03.004},
  DOI = {10.1016/j.physrep.2012.03.004},
  number = {3–4},
  journal = {Physics Reports},
  publisher = {Elsevier BV},
  author = {Vicsek,  Tamás and Zafeiris,  Anna},
  year = {2012},
  month = aug,
  pages = {71–140}
}

@article{Janosov2017,
  title = {Group chasing tactics: how to catch a faster prey},
  volume = {19},
  ISSN = {1367-2630},
  url = {http://dx.doi.org/10.1088/1367-2630/aa69e7},
  DOI = {10.1088/1367-2630/aa69e7},
  number = {5},
  journal = {New Journal of Physics},
  publisher = {IOP Publishing},
  author = {Janosov,  Milán and Virágh,  Csaba and Vásárhelyi,  Gábor and Vicsek,  Tamás},
  year = {2017},
  month = may,
  pages = {053003}
}

@article{Ohira2024,
  title = {Collective behaviors emerging from chases and escapes},
  volume = {29},
  ISSN = {1614-7456},
  url = {http://dx.doi.org/10.1007/s10015-023-00928-1},
  DOI = {10.1007/s10015-023-00928-1},
  number = {1},
  journal = {Artificial Life and Robotics},
  publisher = {Springer Science and Business Media LLC},
  author = {Ohira,  Toru},
  year = {2024},
  month = jan,
  pages = {1–11}
}

@article{Bechinger2016PMR,
  title = {Active particles in complex and crowded environments},
  author = {Bechinger, Clemens and Di Leonardo, Roberto and L\"owen, Hartmut and Reichhardt, Charles and Volpe, Giorgio and Volpe, Giovanni},
  journal = {Rev. Mod. Phys.},
  volume = {88},
  issue = {4},
  pages = {045006},
  numpages = {50},
  year = {2016},
  month = {Nov},
  publisher = {American Physical Society},
  doi = {10.1103/RevModPhys.88.045006},
  url = {https://link.aps.org/doi/10.1103/RevModPhys.88.045006}
}

@article{Bowick22PRX,
  title = {Symmetry, Thermodynamics, and Topology in Active Matter},
  author = {Bowick, Mark J. and Fakhri, Nikta and Marchetti, M. Cristina and Ramaswamy, Sriram},
  journal = {Phys. Rev. X},
  volume = {12},
  issue = {1},
  pages = {010501},
  numpages = {27},
  year = {2022},
  month = {Feb},
  publisher = {American Physical Society},
  doi = {10.1103/PhysRevX.12.010501},
  url = {https://link.aps.org/doi/10.1103/PhysRevX.12.010501}
}

@article{Epstein2019JCP,
  title = {Statistical mechanics of transport processes in active fluids. II. Equations of hydrodynamics for active Brownian particles},
  volume = {150},
  ISSN = {1089-7690},
  url = {http://dx.doi.org/10.1063/1.5054912},
  DOI = {10.1063/1.5054912},
  number = {16},
  journal = {The Journal of Chemical Physics},
  publisher = {AIP Publishing},
  author = {Epstein,  Jeffrey M. and Klymko,  Katherine and Mandadapu,  Kranthi K.},
  year = {2019},
  month = apr 
}

@article{teVrugt2025EPJE,
  title = {Metareview: a survey of active matter reviews},
  volume = {48},
  ISSN = {1292-895X},
  url = {http://dx.doi.org/10.1140/epje/s10189-024-00466-z},
  DOI = {10.1140/epje/s10189-024-00466-z},
  number = {2–3},
  journal = {The European Physical Journal E},
  publisher = {Springer Science and Business Media LLC},
  author = {te Vrugt,  Michael and Wittkowski,  Raphael},
  year = {2025},
  month = mar 
}

@article{evans1993RMP,
  title={Random and cooperative sequential adsorption},
  author={Evans, James W},
  journal={Reviews of modern physics},
  volume={65},
  number={4},
  pages={1281},
  year={1993},
  publisher={APS},
  doi = {https://doi.org/10.1103/RevModPhys.65.1281}
}

@article{huang2024MDPI,
  title={Chaotic-Based Shellcode Encryption: A New Strategy for Bypassing Antivirus Mechanisms},
  author={Huang, Gang-Cheng and Chang, Ko-Chin and Lai, Tai-Hung},
  journal={Symmetry},
  volume={16},
  number={11},
  pages={1526},
  year={2024},
  publisher={MDPI},
  doi = {10.3390/sym16111526}
}

@article{brighton2023CB,
  title={Obstacle avoidance in aerial pursuit},
  author={Brighton, Caroline H and Kempton, James A and France, Lydia A and KleinHeerenbrink, Marco and Mi{\~n}ano, Sof{\'\i}a and Taylor, Graham K},
  journal={Current Biology},
  volume={33},
  number={15},
  pages={3192--3202},
  year={2023},
  publisher={Elsevier},
  doi = {10.1016/j.cub.2023.06.047}
}

@inproceedings{guo2026physpatch,
  title={Physpatch: A physically realizable and transferable adversarial patch attack for multimodal large language models-based autonomous driving systems},
  author={Guo, Qi and Jia, Xiaojun and Pang, Shanmin and Qin, Simeng and Wang, Lin and Jia, Ju and Liu, Yang and Guo, Qing},
  booktitle={Proceedings of the AAAI Conference on Artificial Intelligence},
  volume={40},
  number={6},
  pages={4412--4420},
  year={2026},
  doi = {https://doi.org/10.1609/aaai.v40i6.42439}
}

@article{deng2025PR,
  title={Targeted attack via adversarial patch outside bounding box},
  author={Deng, Kang and Chen, Qixiang and Zhang, Yu and Lin, Zhi and Gong, Shenjian and Liang, Zhenyu and Peng, Anjie and Yang, Xing and Lian, Defu},
  journal={Pattern Recognition},
  pages={112244},
  year={2025},
  publisher={Elsevier},
  doi = {10.1016/j.patcog.2025.112244}
}

@article{Fowell2021,
  title = {The spatio-temporal control of effector T cell migration},
  volume = {21},
  ISSN = {1474-1741},
  url = {http://dx.doi.org/10.1038/s41577-021-00507-0},
  DOI = {10.1038/s41577-021-00507-0},
  number = {9},
  journal = {Nature Reviews Immunology},
  publisher = {Springer Science and Business Media LLC},
  author = {Fowell,  Deborah J. and Kim,  Minsoo},
  year = {2021},
  month = feb,
  pages = {582–596}
}

@article{Schienstock2021,
  title = {Moving beyond velocity: Opportunities and challenges to quantify immune cell behavior*},
  volume = {306},
  ISSN = {1600-065X},
  url = {http://dx.doi.org/10.1111/imr.13038},
  DOI = {10.1111/imr.13038},
  number = {1},
  journal = {Immunological Reviews},
  publisher = {Wiley},
  author = {Schienstock,  Dominik and Mueller,  Scott N.},
  year = {2021},
  month = nov,
  pages = {123–136}
}

@article{MoreiraSoares2020,
  title = {Adhesion modulates cell morphology and migration within dense fibrous networks},
  volume = {32},
  ISSN = {1361-648X},
  url = {http://dx.doi.org/10.1088/1361-648X/ab7c17},
  DOI = {10.1088/1361-648x/ab7c17},
  number = {31},
  journal = {Journal of Physics: Condensed Matter},
  publisher = {IOP Publishing},
  author = {Moreira-Soares,  Maurício and Cunha,  Susana P and Bordin,  José Rafael and Travasso,  Rui D M},
  year = {2020},
  month = may,
  pages = {314001}
}

@article{Melo2023,
  title = {The ECM and tissue architecture are major determinants of early invasion mediated by E-cadherin dysfunction},
  volume = {6},
  ISSN = {2399-3642},
  url = {http://dx.doi.org/10.1038/s42003-023-05482-x},
  DOI = {10.1038/s42003-023-05482-x},
  number = {1},
  journal = {Communications Biology},
  publisher = {Springer Science and Business Media LLC},
  author = {Melo,  Soraia and Guerrero,  Pilar and Moreira Soares,  Maurício and Bordin,  José Rafael and Carneiro,  Fátima and Carneiro,  Patrícia and Dias,  Maria Beatriz and Carvalho,  João and Figueiredo,  Joana and Seruca,  Raquel and Travasso,  Rui D. M.},
  year = {2023},
  month = nov 
}

@article{Gorini11,
  title = {Habitat heterogeneity and mammalian predator–prey interactions},
  volume = {42},
  ISSN = {1365-2907},
  url = {http://dx.doi.org/10.1111/j.1365-2907.2011.00189.x},
  DOI = {10.1111/j.1365-2907.2011.00189.x},
  number = {1},
  journal = {Mammal Review},
  publisher = {Wiley},
  author = {GORINI,  Lucrezia and LINNELL,  John D. C. and MAY,  Roel and PANZACCHI,  Manuela and BOITANI,  Luigi and ODDEN,  Morten and NILSEN,  Erlend. B.},
  year = {2011},
  month = apr,
  pages = {55–77}
}

@article{Cozzi2013,
  title = {Comparison of the effects of artificial and natural barriers on large African carnivores: Implications for interspecific relationships and connectivity},
  volume = {82},
  ISSN = {1365-2656},
  url = {http://dx.doi.org/10.1111/1365-2656.12039},
  DOI = {10.1111/1365-2656.12039},
  number = {3},
  journal = {Journal of Animal Ecology},
  publisher = {Wiley},
  author = {Cozzi,  Gabriele and Broekhuis,  Femke and McNutt,  J. Weldon and Schmid,  Bernhard},
  editor = {Fryxell,  John},
  year = {2013},
  month = feb,
  pages = {707–715}
}

@article{Oyler2016,
  title = {Pursuit–evasion games in the presence of obstacles},
  volume = {65},
  ISSN = {0005-1098},
  url = {http://dx.doi.org/10.1016/j.automatica.2015.11.018},
  DOI = {10.1016/j.automatica.2015.11.018},
  journal = {Automatica},
  publisher = {Elsevier BV},
  author = {Oyler,  Dave W. and Kabamba,  Pierre T. and Girard,  Anouck R.},
  year = {2016},
  month = mar,
  pages = {1–11}
}

@article{Liang2023,
  title = {Collaborative pursuit-evasion game of multi-UAVs based on Apollonius circle in the environment with obstacle},
  volume = {35},
  ISSN = {1360-0494},
  url = {http://dx.doi.org/10.1080/09540091.2023.2168253},
  DOI = {10.1080/09540091.2023.2168253},
  number = {1},
  journal = {Connection Science},
  publisher = {Informa UK Limited},
  author = {Liang,  Xiao and Zhou,  Boran and Jiang,  Linping and Meng,  Guanglei and Xiu,  Yiwei},
  year = {2023},
  month = feb 
}

@article{Katona2024,
  title = {Obstacle Avoidance and Path Planning Methods for Autonomous Navigation of Mobile Robot},
  volume = {24},
  ISSN = {1424-8220},
  url = {http://dx.doi.org/10.3390/s24113573},
  DOI = {10.3390/s24113573},
  number = {11},
  journal = {Sensors},
  publisher = {MDPI AG},
  author = {Katona,  Kornél and Neamah,  Husam A. and Korondi,  Péter},
  year = {2024},
  month = jun,
  pages = {3573}
}

@article{Zhou2021,
  title = {Multi-Target Coordinated Search Algorithm for Swarm Robotics Considering Practical Constraints},
  volume = {15},
  ISSN = {1662-5218},
  url = {http://dx.doi.org/10.3389/fnbot.2021.753052},
  DOI = {10.3389/fnbot.2021.753052},
  journal = {Frontiers in Neurorobotics},
  publisher = {Frontiers Media SA},
  author = {Zhou,  You and Chen,  Anhua and He,  Xinjie and Bian,  Xiaohui},
  year = {2021},
  month = dec 
}

@article{Yaacoub2021,
  title = {Robotics cyber security: vulnerabilities,  attacks,  countermeasures,  and recommendations},
  volume = {21},
  ISSN = {1615-5270},
  url = {http://dx.doi.org/10.1007/s10207-021-00545-8},
  DOI = {10.1007/s10207-021-00545-8},
  number = {1},
  journal = {International Journal of Information Security},
  publisher = {Springer Science and Business Media LLC},
  author = {Yaacoub,  Jean-Paul A. and Noura,  Hassan N. and Salman,  Ola and Chehab,  Ali},
  year = {2021},
  month = mar,
  pages = {115–158}
}

@article{Alqudsi2025,
  title = {UAV swarms: research,  challenges,  and future directions},
  volume = {72},
  ISSN = {2536-9512},
  url = {http://dx.doi.org/10.1186/s44147-025-00582-3},
  DOI = {10.1186/s44147-025-00582-3},
  number = {1},
  journal = {Journal of Engineering and Applied Science},
  publisher = {Springer Science and Business Media LLC},
  author = {Alqudsi,  Yunes and Makaraci,  Murat},
  year = {2025},
  month = jan 
}

@article{helbing2005TS,
  title={Self-organized pedestrian crowd dynamics: Experiments, simulations, and design solutions},
  author={Helbing, Dirk and Buzna, Lubos and Johansson, Anders and Werner, Torsten},
  journal={Transportation science},
  volume={39},
  number={1},
  pages={1--24},
  year={2005},
  publisher={INFORMS},
  doi = {10.1287/trsc.1040.0108}
  
}

@ARTICLE{Zhangetal2007,
  author  = {Zhang, S. and Wang, R. and Zhang, X.},
  title   = {Identification of overlapping community structure in complex networks using fuzzy c-means clustering},
  journal = {Physica A.}, 
  volume  = {374},
  year    = {2007},
  pages   = {483-490},
  doi = {10.1016/j.physa.2006.07.023}  
}

@article{wang2025JCSS,
  title={Modelling emergent pedestrian evacuation behaviors from intelligent, game-playing agents},
  author={Wang, Yiyu and Ge, Jiaqi and Comber, Alexis},
  journal={Journal of Computational Social Science},
  volume={8},
  number={2},
  pages={49},
  year={2025},
  publisher={Springer},
  doi = {10.1007/s42001-025-00369-9}
}

@article{oropesa2026transport,
  title={Transport properties of active particles moving on adjustable networks},
  author={Oropesa, William GC and de Castro, Pablo and L{\"o}wen, Hartmut and Liarte, Danilo B},
  journal={Soft Matter},
  year={2026},
  publisher={The Royal Society of Chemistry},
  doi = {10.1039/d6sm00135a}
}

@article{berner2023PR,
  title={Adaptive dynamical networks},
  author={Berner, Rico and Gross, Thilo and Kuehn, Christian and Kurths, J{\"u}rgen and Yanchuk, Serhiy},
  journal={Physics Reports},
  volume={1031},
  pages={1--59},
  year={2023},
  publisher={Elsevier},
  doi = {j.physrep.2023.08.001}
}

@article{fialkowski2023PRL,
  title={Heterogeneous nucleation in finite-size adaptive dynamical networks},
  author={Fialkowski, Jan and Yanchuk, Serhiy and Sokolov, Igor M and Sch{\"o}ll, Eckehard and Gottwald, Georg A and Berner, Rico},
  journal={Physical Review Letters},
  volume={130},
  number={6},
  pages={067402},
  year={2023},
  publisher={APS},
  doi = {10.1103/PhysRevLett.130.067402}
}

@article{sawicki2023AIP,
  title={Perspectives on adaptive dynamical systems},
  author={Sawicki, Jakub and Berner, Rico and Loos, Sarah AM and Anvari, Mehrnaz and Bader, Rolf and Barfuss, Wolfram and Botta, Nicola and Brede, Nuria and Franovi{\'c}, Igor and Gauthier, Daniel J and others},
  journal={Chaos: An Interdisciplinary Journal of Nonlinear Science},
  volume={33},
  number={7},
  year={2023},
  publisher={AIP Publishing},
  doi = {10.1063/5.0147231}
}

@article{chepizhko13,
  title = {Diffusion, Subdiffusion, and Trapping of Active Particles in Heterogeneous Media},
  author = {Chepizhko, Oleksandr and Peruani, Fernando},
  journal = {Phys. Rev. Lett.},
  volume = {111},
  issue = {16},
  pages = {160604},
  numpages = {5},
  year = {2013},
  month = {Oct},
  publisher = {American Physical Society},
  doi = {10.1103/PhysRevLett.111.160604},
  url = {https://link.aps.org/doi/10.1103/PhysRevLett.111.160604}
}

@article{Bellomo2020,
  title = {Active particles methods and challenges in behavioral systems},
  volume = {30},
  ISSN = {1793-6314},
  url = {http://dx.doi.org/10.1142/S0218202520020017},
  DOI = {10.1142/s0218202520020017},
  number = {04},
  journal = {Mathematical Models and Methods in Applied Sciences},
  publisher = {World Scientific Pub Co Pte Ltd},
  author = {Bellomo,  N. and Brezzi,  F. and Soler,  J.},
  year = {2020},
  month = mar,
  pages = {653–658}
}

@article{rossatto2026confinement,
  title = {Confinement-controlled chase–escape dynamics},
  volume = {692},
  ISSN = {0378-4371},
  url = {http://dx.doi.org/10.1016/j.physa.2026.131524},
  DOI = {10.1016/j.physa.2026.131524},
  journal = {Physica A: Statistical Mechanics and its Applications},
  publisher = {Elsevier BV},
  author = {Rossatto,  R.G. and Alvarez,  H. Ariel and Carlevaro,  C.M. and Bordin,  José Rafael},
  year = {2026},
  month = June,
  pages = {131524}
}

@article{vscepanovic2019PhysicaA,
  title={Group chase and escape in the presence of obstacles},
  author={{\v{S}}{\'c}epanovi{\'c}, JR and Kara{\v{c}}, A and Jak{\v{s}}i{\'c}, ZM and Budinski-Petkovi{\'c}, Lj and Vrhovac, SB},
  journal={Physica A: Statistical Mechanics and its Applications},
  volume={525},
  pages={450--465},
  year={2019},
  publisher={Elsevier},
  doi = {10.1016/j.physa.2019.03.017}
}

@article{vymetal2011ACS,
  title={Gyration-and inertia-tensor-based collective coordinates for metadynamics. Application on the conformational behavior of polyalanine peptides and Trp-cage folding},
  author={Vymetal, Jir{\'\i} and Vondrasek, Jirri{\'\i}},
  journal={The Journal of Physical Chemistry A},
  volume={115},
  number={41},
  pages={11455--11465},
  year={2011},
  publisher={ACS Publications},
  doi = {https://doi.org/10.1021/jp2065612}
  
}

@article{anila2025MDPI,
  title={Theoretical Methods for Assessing the Density of Protein Nanodroplets},
  author={Anila, Midhun Mohan and Wojciechowski, Micha{\l} and Chwastyk, Mateusz and R{\'o}{\.z}ycki, Bartosz},
  journal={International Journal of Molecular Sciences},
  volume={26},
  number={17},
  pages={8631},
  year={2025},
  publisher={MDPI},
  doi = {10.3390/ijms26178631}
}

@article{Demeuldre2017,
    author = {Demeuldre, Mélanie and Hennebert, Elise and Bonneel, Marie and Lengerer, Birgit and Van Dyck, Séverine and Wattiez, Ruddy and Ladurner, Peter and Flammang, Patrick},
    title = {Mechanical adaptability of sea cucumber Cuvierian tubules involves a mutable collagenous tissue},
    journal = {Journal of Experimental Biology},
    volume = {220},
    number = {11},
    pages = {2108-2119},
    year = {2017},
    month = {06},
    issn = {0022-0949},
    doi = {10.1242/jeb.145706},
    url = {https://doi.org/10.1242/jeb.145706},
    eprint = {https://journals.biologists.com/jeb/article-pdf/220/11/2108/1895529/jeb145706.pdf},
}

@article{Flammang2002,
    author = {Flammang, Patrick and Ribesse, Jérôme and Jangoux, Michel},
    title = {Biomechanics of Adhesion in Sea Cucumber Cuvierian Tubules (Echinodermata, Holothuroidea)},
    journal = {Integrative and Comparative Biology},
    volume = {42},
    number = {6},
    pages = {1107-1115},
    year = {2002},
    month = {12},
    issn = {1540-7063},
    doi = {10.1093/icb/42.6.1107},
    url = {https://doi.org/10.1093/icb/42.6.1107},
    eprint = {https://academic.oup.com/icb/article-pdf/42/6/1107/2055212/i1540-7063-042-06-1107.pdf},
}

@article{masuko2017effect,
  title={The effect of laziness in group chase and escape},
  author={Masuko, Makoto and Hiraoka, Takayuki and Ito, Nobuyasu and Shimada, Takashi},
  journal={Journal of the Physical Society of Japan},
  volume={86},
  number={8},
  pages={085002},
  year={2017},
  publisher={The Physical Society of Japan},
  doi = {10.7566/JPSJ.86.085002}
}

@article{somemori2025optimal,
  title={Optimal group structure for group chase and escape},
  author={Somemori, Kohsuke and Shimada, Takashi},
  journal={Artificial Life and Robotics},
  volume={30},
  number={1},
  pages={143--147},
  year={2025},
  publisher={Springer},
  doi = {10.1007/s10015-024-00991-2}
}

@article{vandenspiegel1987,
  title={Cuvierian tubules of the holothuroid Holothuria forskali (Echinodermata): a morphofunctional study},
  author={VandenSpiegel, Didier and Jangoux, Michel},
  journal={Marine Biology},
  volume={96},
  number={2},
  pages={263--275},
  year={1987},
  publisher={Springer}, 
  doi = {10.1007/BF00427026}
}

@article{Matz2004,
author = {Matz, Carsten and Bergfeld, Tanja and Rice, Scott A. and Kjelleberg, Staffan},
title = {Microcolonies, quorum sensing and cytotoxicity determine the survival of Pseudomonas aeruginosa biofilms exposed to protozoan grazing},
journal = {Environmental Microbiology},
volume = {6},
number = {3},
pages = {218-226},
doi = {https://doi.org/10.1111/j.1462-2920.2004.00556.x},
url = {https://enviromicro-journals.onlinelibrary.wiley.com/doi/abs/10.1111/j.1462-2920.2004.00556.x},
eprint = {https://enviromicro-journals.onlinelibrary.wiley.com/doi/pdf/10.1111/j.1462-2920.2004.00556.x},
year = {2004}
}

@article{Weitere2005,
author = {Weitere, Markus and Bergfeld, Tanja and Rice, Scott A. and Matz, Carsten and Kjelleberg, Staffan},
title = {Grazing resistance of Pseudomonas aeruginosa biofilms depends on type of protective mechanism, developmental stage and protozoan feeding mode},
journal = {Environmental Microbiology},
volume = {7},
number = {10},
pages = {1593-1601},
doi = {https://doi.org/10.1111/j.1462-2920.2005.00851.x},
url = {https://enviromicro-journals.onlinelibrary.wiley.com/doi/abs/10.1111/j.1462-2920.2005.00851.x},
eprint = {https://enviromicro-journals.onlinelibrary.wiley.com/doi/pdf/10.1111/j.1462-2920.2005.00851.x},
year = {2005}
}

@article{EscobarAgudelo2026,
  title = {Effective-medium theory for elastic systems with correlated disorder},
  volume = {8},
  ISSN = {2643-1564},
  url = {http://dx.doi.org/10.1103/pdkn-c3z6},
  DOI = {10.1103/pdkn-c3z6},
  number = {2},
  journal = {Physical Review Research},
  publisher = {American Physical Society (APS)},
  author = {Escobar-Agudelo,  Jorge M. and Aquino,  Rui and Liarte,  Danilo B.},
  year = {2026},
  pages={023099},
  month = Apr 
}

@article{Caprini2025,
  title = {Spontaneous self-wrapping in chiral active polymers},
  volume = {1},
  ISSN = {2950-6360},
  url = {http://dx.doi.org/10.1016/j.newton.2025.100253},
  DOI = {10.1016/j.newton.2025.100253},
  number = {9},
  journal = {Newton},
  publisher = {Elsevier BV},
  author = {Caprini,  Lorenzo and Abdoli,  Iman and Marini Bettolo Marconi,  Umberto and L\"{o}wen,  Hartmut},
  year = {2025},
  month = Nov,
  pages = {100253}
}

@article{Vasilyev2017,
  title = {Cooperative behavior of biased probes in crowded interacting systems},
  volume = {13},
  ISSN = {1744-6848},
  url = {http://dx.doi.org/10.1039/c7sm00865a},
  DOI = {10.1039/c7sm00865a},
  number = {41},
  journal = {Soft Matter},
  publisher = {Royal Society of Chemistry (RSC)},
  author = {Vasilyev,  Oleg A. and Bénichou,  Olivier and Mejía-Monasterio,  Carlos and Weeks,  Eric R. and Oshanin,  Gleb},
  year = {2017},
  pages = {7617–7624}
}

@article{Reichhardt2006,
  title = {Cooperative behavior and pattern formation in mixtures of driven and nondriven colloidal assemblies},
  volume = {74},
  ISSN = {1550-2376},
  url = {http://dx.doi.org/10.1103/PhysRevE.74.011403},
  DOI = {10.1103/physreve.74.011403},
  number = {1},
  journal = {Physical Review E},
  publisher = {American Physical Society (APS)},
  author = {Reichhardt,  C. and Reichhardt,  C. J. Olson},
  year = {2006},
  month = July 
}

\end{document}